# Engineering electron and hole tunneling with asymmetric InAs quantum dot molecules

A. S. Bracker, M. Scheibner, M. F. Doty, E. A. Stinaff, I. V. Ponomarev,

J. C. Kim, L. J. Whitman, T. L. Reinecke, D. Gammon

*Naval Research Laboratory, Washington DC 20375, USA*

Most self-assembled quantum dot molecules are intrinsically asymmetric with inequivalent dots resulting from imperfect control of crystal growth. We have grown vertically-aligned pairs of InAs/GaAs quantum dots by molecular beam epitaxy, introducing intentional asymmetry that limits the influence of intrinsic growth fluctuations and allows selective tunneling of electrons or holes. We present a systemic investigation of tunneling energies over a wide range of interdot barrier thickness. The concepts discussed here provide an important tool for the systematic design and characterization of more complicated quantum dot nanostructures.

Quantum mechanical tunneling of electrons and holes between self-assembled semiconductor quantum dots creates "molecular" states of great technological interest. When combined with exchange interactions, tunneling allows carrier spins to be manipulated by optical[1,2] or electrostatic[3,4] fields, thereby providing a potential entanglement mechanism for quantum information technology.

Early spectroscopic measurements on nominally symmetric quantum dot molecules (QDMs) reported carrier tunneling,[5,6] but experimental progress in the field slowed for several years following those pioneering studies. The main catalyst for renewed activity in recent months has been the direct observation of coherent tunneling in QDMs embedded in electric field-tunable Schottky diodes.[1,2] Coherence is observed clearly as "anticrossings" in a two-dimensional plot of the photoluminescence (PL) spectrum vs. electric field. The anticrossings appear where the optical transitions for *intra*dot and *inter*dot excitons meet. Interdot excitons (electron and hole in different QDs) have a large Stark shift, i.e. their energies vary rapidly with electric field, while intradot excitons show a weak Stark shift. At the anticrossing, the orbital wavefunctions of the exciton take on "bonding" and "antibonding" character.

The first measurements of anticrossings in our laboratory were on nominally symmetric QDMs, where the individual QD heights were chosen to be the same. These samples gave a perplexing result—we found that the QDMs fell into two distinct categories. Some spectra showed small, sharp anticrossings, while others showed very large anticrossings or even broadly curving lines with no clear upper branch. Below, we prove that this dichotomy arises from natural structural asymmetry in the QDMs. Because of imprecise control of crystal growth, two QDs will have differences in size, shape, and composition that give them distinct transition energies. Depending on whether the exciton energy is larger for the top or bottom dot in a QDM, we observe tunneling of electrons or holes individually, rather than simultaneously as an exciton. The two cases appear distinctly different, because holes have a larger effective mass and therefore a lower tunneling rate, which results in smaller anticrossing energies than electrons. Unintended growth asymmetry therefore explains why both hole and electron tunneling have been observed recently in nominally-symmetric QDMs.[1,2,7] Even for the ideal case of two dots with perfectly equivalent structures, an intrinsic lack of reflection symmetry leads to energetic inequivalence.[8,9] For both practical and fundamental reasons, therefore, asymmetry should be viewed not as a flaw, but as an essential design choice that provides an opportunity to take advantage of the very different properties of electrons and holes.

Controlling asymmetry requires changes to typical self-assembly techniques. Self-assembled InAs QDs grown by molecular beam epitaxy (MBE) on a GaAs surface have the shape of facetted domes or truncated pyramids. However, the QD shapes and dimensions, as well as the effects of subsequent overgrowth with GaAs, are highly sensitive to growth conditions. Furthermore, in a vertically-stacked QDM, the upper dot is usually larger than the bottom dot, because of strain-enhanced nucleation. In our work, the key to controlling the QDM asymmetry is the height of the individual QDs. Height control is obtained with the "indium flush" growth method,[10] where the as-grown QD is partially capped[11] with GaAs and then annealed at a higher temperature. This growth sequence effectively shears the top off of the as-grown dot, producing a disk-shaped QD with a top surface that is roughly coplanar with the GaAs capping layer surface. The height of the GaAs cap can be controlled with monolayer accuracy and largely determines the QD height, which has a major influence on the confinement energy of the QD.

In a simple experiment, we prove that moderate asymmetries produce two qualitatively different types of QDMs. We compare two samples with a large degree of intentional asymmetry [schematics in Fig. 1(a)], grown using the indium flush technique. InAs QDs were deposited on a GaAs buffer layer at 520°C, and a GaAs partial cap was grown. The sample was annealed for 70 seconds at 570°C to truncate the QD height, and after growing a thin GaAs interdot barrier, the procedure was repeated for the second QD. In the first sample, the heights were 4.0 nm on the top (T) and 2.5 nm on bottom (B), while in the second sample, the order of the two QD sizes was reversed. Cross-sectional scanning tunneling microscopy images[12] of these two types of QDMs are shown in Fig. 1(b). For low temperature PL spectroscopy measurements, the QDMs were embedded in an n-I Schottky diode in order to control the electric field. Because of the built-in electric field near a GaAs device surface, this heterostructure has a positive electric field pointing in the sample growth direction, and this field can be changed with an applied bias.

The band edge diagrams of Fig. 1(c) show how the QDM asymmetry determines which type of carrier tunnels. If the top QD is thicker, it has a smaller exciton energy, so a positive electric field across the QDM brings the individual QD electron levels into resonance, while the hole levels are detuned. When the order of the QDs is reversed, the positive electric field brings the QD hole levels into resonance. These situations can be seen clearly in calculated exciton energy diagrams for QDMs with opposite asymmetries [Figs 2(a) and 2(b)] (see also, Ref. 13.) We focus on the anticrossings highlighted by yellow circles in the energy levels of the larger (lower energy) QD, which are observed in a PL experiment. In the first type of sample [Fig. 2(a)], an electron tunneling resonance with a large anticrossing occurs at a positive electric field where the intradot exciton $^{01}_{01}X^0$ of the lower energy (top) QD crosses the interdot transition

$^{10}_{01}X^0$. With the opposite QDM asymmetry in the second sample, a hole tunneling resonance occurs at positive electric field, and the anticrossing is much smaller [Fig. 2(b)].

In the PL spectra, we observe only small anticrossings in the sample designed for hole tunneling and only large anticrossings in the sample designed for electron tunneling. Examples are shown in Figs 3(a) and 3(b), corresponding to measurements in the region of positive electric field in Figs 2(a) and 2(b), respectively. This result shows clearly that by selecting the order of the dots in a QDM sample, with all else kept the same, we can select whether electrons or holes tunnel. The individual QDs with heights of 2.5 nm and 4 nm have intradot exciton energies that differ by around 90 meV, which is more than twice as large as the typical inhomogeneous energy broadening caused by intrinsic structural variations. This guarantees that the large majority of QDMs will have the desired energy ordering and thereby the desired type of carrier tunneling. With this approach, we avoid the ambiguities brought about by growth fluctuations and can directly access the fundamental physics revealed by the optical spectra.

An alternative method of selecting between electron and hole tunneling is through the sign of the electric field. This alternative can be seen by comparing the anticrossings on the left and the right of the exciton energy diagrams in Fig. 2(a) or 2(b). For example, a sample with the appropriate asymmetry [Fig. 2(a)] to give electron tunneling with a positive electric field would give hole tunneling with a negative field. In our n-I Schottky diodes, it is not practical to reverse the sign of the electric field with an applied bias, because this would flood the structure with electrons. However, a straightforward alternative is to use a p-I Schottky diode, which has a negative built-in field.

With the ability to select between electron or hole tunneling, we systematically examine the influence of barrier thickness. We use QDMs with the thinner QD on top, and a p-I or n-I

Schottky diode to select electron or hole tunneling, respectively. For each QDM, we measured the single electron ($\Delta_e$) or single hole ($\Delta_h$) anticrossing energies in the spectral 'x'-patterns of negative and positive trions. Fig. 4 shows data points corresponding to measurements on 62 QDMs in twelve samples, with linear fits through the data points. The electron anticrossing energies are more than ten times larger than the hole values within the measured range, and the slopes of the data sets differ by roughly a factor of two. Both effects arise in part from the higher effective mass of holes.[14]

We observe a large scatter in the anticrossing energies for each sample. A typical case (hole anticrossing, 6 nm barrier) is shown in the inset of Fig. 4. The most obvious explanation for this scatter would be variations in the barrier thickness for different QDMs. However, the Stark shifts of the interdot PL transitions provide a measure of the interdot separation, and we observe no correlation between anticrossing energy and interdot Stark shifts for a given sample. This is not surprising, because the indium flush technique is expected to give accurate control over the interdot barrier thickness. Instead, the spread in anticrossing energies must arise from variations in other QDM properties such as the lateral size of the individual QDs, lateral alignment between the QDs, alloy composition, and other complex three-dimensional structure. These properties are sensitive to MBE growth conditions, and we believe it is likely that similar samples from different laboratories will show considerable variations in tunneling probability. Techniques to improve QD homogeneity[15] will further enhance the technological promise of self-assembled QDMs.

The results presented here suggest a natural step forward in controlling tunneling within larger quantum dot complexes. Because the relative vertical heights of two QDs can be controlled reproducibly through crystal growth, it will be possible to specify the sequence of relative energies within a larger chain of QDs, at least for nearest neighbor pairs. As the ability to laterally position dots improves, three-dimensional networks of coupled quantum dots will become feasible. With further development in optical spin manipulation through exchange interactions, we anticipate that these systems will serve as prototypes for simple multi-qubit manipulations.

This work was supported by NSA/ARO and ONR. MFD, EAS, IVP, and JCK are NRL/NRC Research Associates.

Figure Captions

Fig. 1

(a) Asymmetric QDM structures designed for electron or hole tunneling at positive electric fields. (b) Cross-sectional scanning tunneling microscopy images of both types of QDM. (c) Band edge potentials for both types of QDM, resulting in electron and hole tunneling resonances, respectively.

Fig. 2

(a) Calculation of exciton spectrum for first type of QDM from Fig. 1. Higher and lower energy horizontal lines correspond to intradot exciton energies for bottom dot and top dot, respectively. Sloped lines are interdot exciton energies. Exciton symbols are defined by $_{hB,hT}^{eB,eT}X^0$. Electron tunneling resonance occurs at positive electric field, while hole resonance occurs at negative field. (b) Calculated exciton spectrum for second type of QDM. The energy ordering of the bottom and top dots is reversed, which reverses the electric field ordering of electron and hole resonances.

Fig. 3

(a) PL intensity as a function of PL energy and electric field for the first type of QDM in Fig. 1. (b) Same for second type of QDM in Fig. 1. Electric field and PL energy scales have equal proportions in both graphs. Other features in these graphs result from biexcitons and charged excitons.[2]

Fig. 4

QDM anticrossing energies of a single electron ($\Delta_e$) or a single hole ($\Delta_h$) as a function of barrier thickness, taken from charged exciton PL spectra (circles). Squares indicate anticrossings measured from neutral excitons. Error bars on solid points indicate standard deviations calculated from measurements on multiple QDMs. Hollow points indicate individual measurements. Lines are linear fits (on a semilog scale) to all of the data points. Inset shows distribution of hole anticrossing energies for all data points in the d = 6 nm sample.

Fig. 1

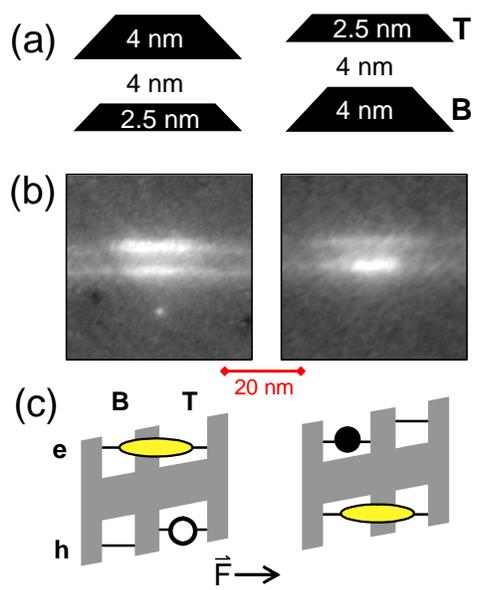

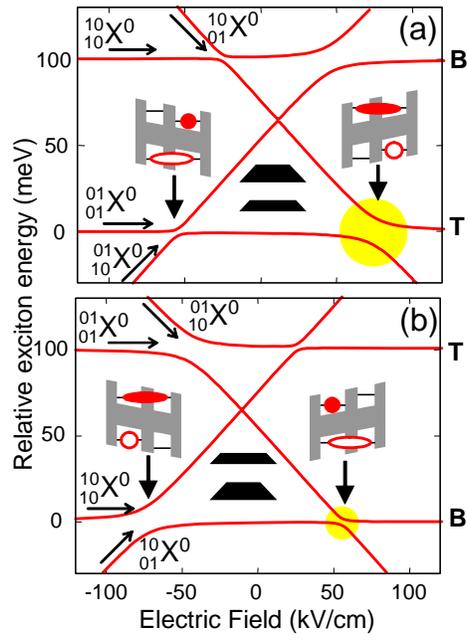

Fig. 3

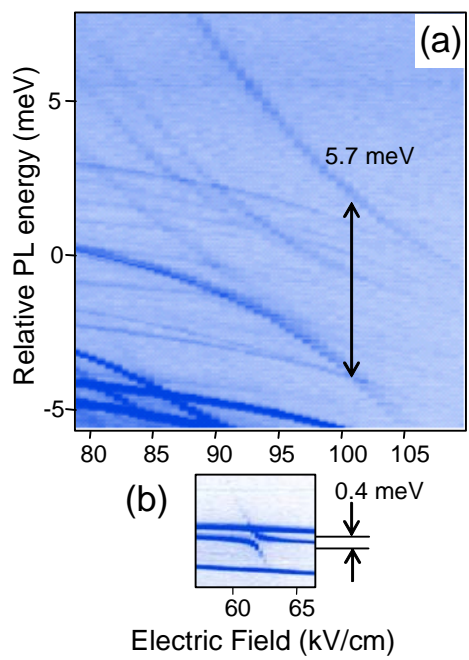

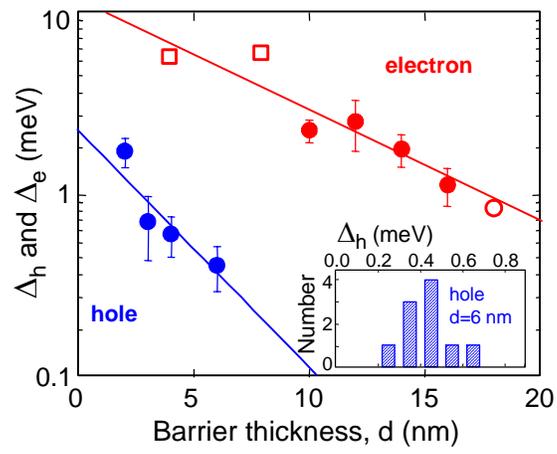